\documentclass[a4paper,aps,twocolumn,nofootinbib]{revtex4}
\RequirePackage[english]{babel}
\RequirePackage[latin1]{inputenc}
\RequirePackage[T1]{fontenc}
\RequirePackage{mathrsfs}
\RequirePackage{amsmath}
\RequirePackage{amssymb}
\RequirePackage{amsbsy}
\RequirePackage{bm}
\begin{document}
%%%%%%%%%%%%%%%%%%%%%%%%%%%%%%%%%%%%%%%%%%%%%%%%%%%%%%%%%%%%%%%%%%%%%%%%%%%%%%%%%%%%%%%%%%%%%%%%%%%
\title{\bf{A simple assessment on the hierarchy problem}}
\author{Luca Fabbri}
\affiliation{DIME Sez. Metodi e Modelli Matematici, Universit\`{a} di Genova,\\
Piazzale Kennedy Pad.D, 16129 Genova, Italy}
\date{\today}
%%%%%%%%%%%%%%%%%%%%%%%%%%%%%%%%%%%%%%%%%%%%%%%%%%%%%%%%%%%%%%%%%%%%%%%%%%%%%%%%%%%%%%%%%%%%%%%%%%%
\begin{abstract}
We consider the simplest extension of the standard model, where torsion couples to spinor as well as to scalar fields, and in which the cosmological constant problem is solved.
\end{abstract}
%%%%%%%%%%%%%%%%%%%%%%%%%%%%%%%%%%%%%%%%%%%%%%%%%%%%%%%%%%%%%%%%%%%%%%%%%%%%%%%%%%%%%%%%%%%%%%%%%%%
\maketitle
%%%%%%%%%%%%%%%%%%%%%%%%%%%%%%%%%%%%%%%%%%%%%%%%%%%%%%%%%%%%%%%%%%%%%%%%%%%%%%%%%%%%%%%%%%%%%%%%%%%
\section{Introduction}
The problem of the incompatibility between the cosmological constant and the Higgs field is well known, and it comes from the fact that whenever the Higgs boson acquires the vacuum expectation value which minimizes its potential, the minimum of the potential gives rise to an effective cosmological constant that is negative and one-hundred and twenty orders of magnitude off the empirically measured value: to solve this problem one may look for another mechanism producing an effective cosmological constant with positive value and a fine-tuning up to one-hundred and twenty decimal places yielding a positive small cosmological constant; otherwise, the only possibility is replacing spontaneous symmetry breaking with a dynamical symmetry breaking, with the Higgs being a fermion condensate subject to Nambu--Jona-Lasinio potentials. So where do these NJL potentials come from?

The torsional completion of gravity is the theory we obtain when we do not constrain the most general metric-compatible connection to be symmetric; in a geometry with torsion and curvature applied to Dirac matter fields, torsion couples to the spin density as curvature couples to the energy density, and the resulting theory is called Sciama-Kibble-Einstein-Dirac SKED theory. In it, the Dirac matter field equation has NJL potentials giving the possibility to describe the Higgs as a condensate \cite{Zubkov:2010sx, FabbriCH}.

In this construction however, the Higgs compositeness does not suffice, and physics beyond the standard model is invoked \cite{b-h-l, Dobrescu:1997nm, Chivukula:1998wd, Castillo-Felisola:2013jva}. But this new physics is still unknown.

Then, it may be wiser to look for a simpler model.
%%%%%%%%%%%%%%%%%%%%%%%%%%%%%%%%%%%%%%%%%%%%%%%%%%%%%%%%%%%%%%%%%%%%%%%%%%%%%%%%%%%%%%%%%%%%%%%%%%%
\section{Torsion-Spinor--Scalar Coupling}
In this paper, we take $(1\!+\!3)$-dimensional spacetimes filled with $\frac{1}{2}$-spin spinor fields, and all the fields will be coupled in terms of least-order derivative field equations.

In \cite{Fabbri:2006xq,Fabbri:2009se,Fabbri:2008rq,Fabbri:2009yc,Fabbri:2011kq,Fabbri:2010rw, Fabbri:2012ag,Fabbri:2013gza} one finds the general definitions, and here we recall the fundamental notations. Since we ask the connection $\Gamma^{\alpha}_{\sigma\nu}$ to have a unique symmetric part, the torsion tensor $\Gamma^{\alpha}_{[\sigma\nu]}\!=\!Q^{\alpha}_{\sigma\nu}$ is completely antisymmetric; the completely antisymmetric pseudo-tensor of Levi-Civita has four indices $\varepsilon_{\alpha\sigma\nu\rho}$ as it is very well known: consequently the torsion can be dualized by means of the Levi-Civita pseudo-tensor as $Q_{\alpha\sigma\nu}\varepsilon^{\alpha\sigma\nu\rho}\!=\!W^{\rho}$ as a torsion pseudo-tensor, that is as an axial vector. We introduce the Clifford matrices $\boldsymbol{\gamma}_{\alpha}$ such that $2\boldsymbol{\mathbb{I}}g_{\mu\nu}\!=\!\{\boldsymbol{\gamma}_{\mu},\boldsymbol{\gamma}_{\nu}\}$ in terms of which we also have $\frac{1}{4}[\boldsymbol{\gamma}_{\mu},\boldsymbol{\gamma}_{\nu}]\!=\!\boldsymbol{\sigma}_{\mu\nu}$ so that we can define the $\boldsymbol{\pi}$ matrix as $\{\boldsymbol{\gamma}_{\alpha},\boldsymbol{\sigma}_{\mu\nu}\}\!=\! i\varepsilon_{\alpha\mu\nu\rho} \boldsymbol{\pi}\boldsymbol{\gamma}^{\rho}$ being the parity-odd matrix: with it $\boldsymbol{\pi}L\!=\!-L$ and $\boldsymbol{\pi}R\!=\!R$ are defined to be the left-handed and right-handed chiral projections of the spinor fields. And $\phi$ is the complex scalar field.

To write the dynamics, we neglect gravity, we consider that because torsion can equivalently be written as an axial vector its dynamics is that of a Proca field, and since both Dirac and scalar fields are as usual, the Lagrangian accounting for all renormalizable interactions is given by
\begin{eqnarray}
\nonumber
&L\!=\!\frac{1}{4}(\nabla_{\alpha}W_{\nu}\!-\!\nabla_{\nu}W_{\alpha})
(\nabla^{\alpha}W^{\nu}\!-\!\nabla^{\nu}W^{\alpha})\!-\!\frac{1}{2}M^{2}W^{2}-\\
\nonumber
&-i\overline{L}\boldsymbol{\gamma}^{\mu}\boldsymbol{\nabla}_{\mu}L
\!-\!i\overline{R}\boldsymbol{\gamma}^{\mu}\boldsymbol{\nabla}_{\mu}R
\!-\!\nabla^{\mu}\phi^{\dagger}\nabla_{\mu}\phi+\\
\nonumber
&+YW_{\mu}(\overline{L}\boldsymbol{\gamma}^{\mu}L\!-\!\overline{R}\boldsymbol{\gamma}^{\mu}R)
\!+\!\Xi\phi^{2}W^{2}+\\
&+G(\overline{L}\phi R\!+\!\overline{R}\phi^{\dagger}L)\!+\!\frac{1}{2}\lambda^{2}\phi^{4}
\label{l}
\end{eqnarray}
where $M$ is the mass of torsion with $Y$ and $\Xi$ being constants describing the coupling of torsion to spinor and scalar fields: varying (\ref{l}) with respect to torsion yields
\begin{eqnarray}
\nonumber
&\nabla_{\alpha}(\nabla^{\alpha}W^{\nu}\!-\!\nabla^{\nu}W^{\alpha})\!+\!M^{2}W^{\nu}=\\
&=Y(\overline{L}\boldsymbol{\gamma}^{\nu}L\!-\!\overline{R}\boldsymbol{\gamma}^{\nu}R)
\!+\!2\Xi\phi^{2}W^{\nu}
\end{eqnarray}
for an axial vector with mass; notice that it is from geometrical arguments that we can define the axial vector showing that torsion is not defined in terms of any gauge transformation, and therefore there is nothing that prevents it to be massive even before symmetry breaking.

If before symmetry breaking torsion has a mass large enough to permit a regime in which we may write
\begin{eqnarray}
&M^{2}W^{\nu}\!\approx\!Y(\overline{L}\boldsymbol{\gamma}^{\nu}L
\!-\!\overline{R}\boldsymbol{\gamma}^{\nu}R)\!+\!2\Xi\phi^{2}W^{\nu}
\end{eqnarray}
then torsion can be inverted: substituting this expression into the Lagrangian we obtain the effective Lagrangian
\begin{eqnarray}
\nonumber
&L\!=\!-i\overline{L}\boldsymbol{\gamma}^{\mu}\boldsymbol{\nabla}_{\mu}L
\!-\!i\overline{R}\boldsymbol{\gamma}^{\mu}\boldsymbol{\nabla}_{\mu}R
\!-\!\nabla^{\mu}\phi^{\dagger}\nabla_{\mu}\phi+\\
\nonumber
&+\frac{1}{2}(M^{2}-2\Xi\phi^{2})^{-1}Y^{2}(\overline{L}\boldsymbol{\gamma}L
\!-\!\overline{R}\boldsymbol{\gamma}R)^{2}+\\
&+G(\overline{L}\phi R\!+\!\overline{R}\phi^{\dagger}L)\!+\!\frac{1}{2}\lambda^{2}\phi^{4}
\label{le}
\end{eqnarray}
where the low-energy torsion contribution is an effective interaction between fermions scaled by the scalar field.

In (\ref{le}) the scalar has potential that is minimized when
\begin{eqnarray}
&\lambda^{2}v^{2}(M^{2}\!-\!2\Xi v^{2})^{2}\!+\!\Xi Y^{2}(\overline{L}\boldsymbol{\gamma}L
\!-\!\overline{R}\boldsymbol{\gamma}R)^{2}\!=\!0
\end{eqnarray}
as conditions of stable equilibrium for the vacuum of the scalar given as $\phi^{2}|_{0}\!=\!v^{2}$ so that the general expression in the unitary gauge $\phi^{\dagger}\!=\!(0,v+H)$ gives rise to the
\begin{eqnarray}
\nonumber
&\!\!L\!=\!-i\overline{\nu}\boldsymbol{\gamma}^{\mu}\boldsymbol{\nabla}_{\mu}\nu
\!-\!i\overline{e}\boldsymbol{\gamma}^{\mu}\boldsymbol{\nabla}_{\mu}e
\!+\!Gv\overline{e}e-\\
\nonumber
&\!\!-\nabla^{\mu}H\nabla_{\mu}H\!+\!3\lambda^{2}v^{2}H^{2}-\\
\nonumber
&\!\!-[M^{2}\!-\!2\Xi(v\!+\!H)^{2}]^{-1}Y^{2} 
(\overline{\nu}\boldsymbol{\gamma}^{\mu}\nu\overline{e}\boldsymbol{\gamma}_{\mu}\boldsymbol{\pi}e \!+\!\frac{1}{2}\overline{e}\boldsymbol{\gamma}^{\mu}e\overline{e}\boldsymbol{\gamma}_{\mu}e)+\\
&\!\!+G\overline{e}eH\!+\!\frac{1}{2}\lambda^{2}(H\!+\!4v)H^{3}
\!+\!2\lambda^{2}v^{3}H\!+\!\frac{1}{2}\lambda^{2}v^{4}
\label{leb}
\end{eqnarray}
with $\nu$ neutrino and $e$ electron; this expression is general, but a particularly simplified form can be obtained within the approximation above when also $M^{2}\!\gg\!\Xi v^{2}$ is valid.

If we have the validity of this approximation then the stable equilibrium is obtained when the vacuum of the scalar is $\lambda^{2}v^{2}M^{4}\!=\!2\Xi Y^{2}(\overline{\nu}\boldsymbol{\gamma}^{\mu}\nu\overline{e}\boldsymbol{\gamma}_{\mu}\boldsymbol{\pi}e \!+\!\frac{1}{2}\overline{e}\boldsymbol{\gamma}^{\mu}e\overline{e}\boldsymbol{\gamma}_{\mu}e)$ and in this case the effective Lagrangian (\ref{leb}) breaks down to
\begin{eqnarray}
\nonumber
&L\!=\!-i\overline{\nu}\boldsymbol{\gamma}^{\mu}\boldsymbol{\nabla}_{\mu}\nu
\!-\!i\overline{e}\boldsymbol{\gamma}^{\mu}\boldsymbol{\nabla}_{\mu}e
\!+\!Gv\overline{e}e-\\
&-\nabla^{\mu}H\nabla_{\mu}H\!+\!2\lambda^{2}v^{2}H^{2}
\!-\!\frac{1}{2}\lambda^{2}v^{2}M^{2}\Xi^{-1}\!+\!V_{\mathrm{hd}}
\label{lesm}
\end{eqnarray}
in which $V_{\mathrm{hd}}$ contains all higher-dimension interactions.

From (\ref{lesm}) we may extract the values of the parameters
\begin{eqnarray}
&m_{e}\!=\!Gv\ \ \ \ m_{\mathrm{H}}^{2}\!=\!2\lambda^{2}v^{2}
\ \ \ \ \ \ \ \ \Lambda\!=\!-\frac{\lambda^{2}v^{2}M^{2}}{2\Xi}
\end{eqnarray}
given by the mass of the electron and the mass of the Higgs field together with the cosmological constant.

We notice that with a Fierz rearrangement we can see that $\lambda^{2}v^{2}M^{4}\!=\!\Xi Y^{2}
(2\overline{\nu}\boldsymbol{\gamma}^{\mu}\nu\overline{e}\boldsymbol{\gamma}_{\mu}\boldsymbol{\pi}e
+\!|\overline{e}e|^{2}\!+\!|i\overline{e}\boldsymbol{\pi}e|^{2})$ showing that the constant $\Xi$ must be positive and therefore the cosmological constant $2\Lambda\!=\!-\lambda^{2}v^{2}M^{2}\Xi^{-1}$ is negative as expected in the case of an effective cosmological constant generated by symmetry breaking, although now its value depends also on the torsional constants; the Higgs mass is given as usual and also the electron mass has the form it has in the common version of the standard model.

So leaving aside the electron and Higgs mass, we may focus on the cosmological constant written above in terms of the Higgs vacuum and employing the condition of equilibrium we may rewrite it in terms of the fermion vacuum as $-2\Lambda\!=\!M^{-2}Y^{2}(2\overline{\nu}\boldsymbol{\gamma}^{\mu}\nu\overline{e}\boldsymbol{\gamma}_{\mu}\boldsymbol{\pi}e
\!+\!|\overline{e}e|^{2}\!+\!|i\overline{e}\boldsymbol{\pi}e|^{2})$ showing that the cosmological constant is effectively obtained as the value that is calculated on the vacuum of fermions in the expression that is ultimately given by the fermionic contact spinor-spinor self-interactions which are induced by the presence of the torsion field of the spacetime.

The torsionally-induced fermionic interactions have a minimum that can be large if the fermion density vacuum is large, as in high-energy physics experiments, but such a value may also be small, as in low-energy physics experiments, and certainly negligible when it is interpreted as the cosmological constant, in cosmology.
%%%%%%%%%%%%%%%%%%%%%%%%%%%%%%%%%%%%%%%%%%%%%%%%%%%%%%%%%%%%%%%%%%%%%%%%%%%%%%%%%%%%%%%%%%%%%%%%%%%
\section{Conclusion}
In this paper, we have considered the simplest extension of the standard model, the one given by the torsional completion of gravity, accounting for all possible renormalizable torsional contributions and interactions with the spinor and the scalar fields; we have noticed that there is no symmetry protecting the torsion mass, and so we have studied what happens when this mass is large enough to allow low-energy conditions and to permit the approximation $M^{2}\!\gg\!\Xi v^{2}$ to hold: we have found that the effective Lagrangian after symmetry breaking led to the usual value of the electron and Higgs masses, but the cosmological constant expressed in terms of the scalar vacuum could be re-expressed with the fermion vacuum as $-2\Lambda\!=\!M^{-2}Y^{2}(2\overline{\nu}\boldsymbol{\gamma}^{\mu}\nu\overline{e}\boldsymbol{\gamma}_{\mu}\boldsymbol{\pi}e
\!+\!|\overline{e}e|^{2}\!+\!|i\overline{e}\boldsymbol{\pi}e|^{2})$ hence showing that the cosmological constant is proportional to the vacuum of fermions, with the consequence that for low-density fermions the cosmological constant is small.

The emerging picture is that both cosmological constant and mass generation are due to symmetry breaking through a dynamical scalar whose vacuum depends on the fermion density: within fermions the scalar has non-trivial vacuum and dynamical breakdown of symmetry occurs, but neither cosmological constant nor masses would appear without fermionic distributions.

The in-matter cosmological constant is a gravitational effect impossible to detect, but the vanishing of the cosmological constant in empty space constitutes a solution to the cosmological constant problem.
%%%%%%%%%%%%%%%%%%%%%%%%%%%%%%%%%%%%%%%%%%%%%%%%%%%%%%%%%%%%%%%%%%%%%%%%%%%%%%%%%%%%%%%%%%%%%%%%%%%

%%%%%%%%%%%%%%%%%%%%%%%%%%%%%%%%%%%%%%%%%%%%%%%%%%%%%%%%%%%%%%%%%%%%%%%%%%%%%%%%%%%%%%%%%%%%%%%%%%%
\end{document}